\documentclass[12pt]{article}
\usepackage{latexsym}
\def\bea{\begin{eqnarray}}
\def\eea{\end{eqnarray}}
\def\beq{\begin{equation}}
\def\eeq{\end{equation}}
\newcommand{\gappeq}{\mathrel{\rlap {\raise.5ex\hbox{$>$}}
{\lower.5ex\hbox{$\sim$}}}}
\newcommand{\lappeq}{\mathrel{\rlap{\raise.5ex\hbox{$<$}}
{\lower.5ex\hbox{$\sim$}}}}
\def\pp{\partial}
\def\Tr{{\rm Tr}}
\def\STr{{\rm STr}}
\def\re{{\rm Re}}

\def\mG{m_{\frac{3}{2}}}
\def\mg{m_{\frac{1}{2}}}
\def\m{\bar m}
\def\s{\bar s}
\def\S{\bar S}
\def\T{\bar T}
\def\t{\bar t}
\def\z{\bar z}
\def\half{\frac{1}{2}}
\def\tc{\tilde{c}}
\def\vev{$vev$}
\def\[{\left [}
\def\]{\right ]}
\def\({\left (}
\def\){\right )}
\newcommand{\lang}{\left\langle}
\newcommand{\rang}{\right\rangle}

\newcommand{\jbar}{\bar{\jmath}}

\def\u1{$U(1)$}
\def\ux{$U(1)_X$}
\def\ua{$U(1)_a$}
\def\L{\mathcal{L}}
\def\D{\mathcal{D}}

\begin{document}

\begin{titlepage}

      \hfill  LBNL-56404

      \hfill  UCB-PTH-04/25

      \hfill hep-ph/0410001

\hfill September 2004 \\[.2in]

\begin{center}

{\large \bf Quadratic Divergences in Effective Supergravity from the
Heterotic Superstring}\footnote{Talk presented at the Nathfest,
Northeastern University, Boston, MA, August 14--15, 2004, to be
published in the proceedings of PASCOS04.}\footnote{This work was
supported in part by the Director, Office of Energy Research, Office
of High Energy and Nuclear Physics, Division of High Energy Physics of
the U.S. Department of Energy under Contract DE-AC03-76SF00098 and in
part by the National Science Foundation under grant
PHY-0098840.}\\[.05in]
\vspace{18pt}

Mary K. Gaillard\\[.1in]

{\em Department of Physics, University of California\\ and\\
 Theoretical Physics Group, 50A-5101, Lawrence Berkeley National
 Laboratory\\ Berkeley, CA 94720, USA}\\[.2in]

\end{center}
\vspace{18pt}

\begin{abstract} Results from studies of effective Lagrangians for gaugino
condensation are summarized and re-examined with an eye to previously
neglected one-loop quadratically divergent corrections.
\end{abstract}
\end{titlepage}
\newpage
\renewcommand{\thepage}{\roman{page}}
\setcounter{page}{2}
\mbox{ }

\vskip 1in

\begin{center}
{\bf Disclaimer}
\end{center}

\vskip .2in

\begin{scriptsize}
\begin{quotation}
This document was prepared as an account of work sponsored by the United
States Government.  While this document is believed to contain
correct information, neither the United States Government nor any agency
thereof, nor The Regents of the University of California, nor any of their
employees, makes any warranty, express or implied, or assumes any legal
liability or responsibility for the accuracy, completeness, or usefulness
of any information, apparatus, product, or process disclosed, or represents
that its use would not infringe privately owned rights.  Reference herein
to any specific commercial products process, or service by its trade name,
trademark, manufacturer, or otherwise, does not necessarily constitute or
imply its endorsement, recommendation, or favoring by the United States
Government or any agency thereof, or The Regents of the University of
California.  The views and opinions of authors expressed herein do not
necessarily state or reflect those of the United States Government or any
agency thereof, or The Regents of the University of California.
\end{quotation}
\vfill

\end{scriptsize}

\vskip 2in

\begin{center}
\begin{small}
{\it Lawrence Berkeley Laboratory is an equal opportunity employer.}
\end{small}
\end{center}

\newpage
\renewcommand{\theequation}{\arabic{section}.\arabic{equation}}
\renewcommand{\thepage}{\arabic{page}}
\setcounter{page}{1}
\def\thefootnote{\arabic{footnote}}
\setcounter{footnote}{0}

\section{Introduction}
The subject of my talk is closely related to the pioneering
work~\cite{can} of Arnowitt, Chamseddine and Nath on locally
supersymmetric Grand Unified Theories, with the difference that the
theory above the scale of unification is a string theory rather than a
field theory.  Specifically, I will consider effective supergravity
theories obtained from compactification of the weakly coupled
heterotic string.  In the next section I summarize
results~\cite{bgw,ggm} from the study of modular (T-duality) invariant
effective Lagrangians for gaugino condensation.  These are
characterized in particular by
\begin{itemize}
\item Dilaton dominated supersymmetry breaking. The auxiliary fields
of the T-moduli (or K\"ahler moduli) have vanishing vacuum values
(\vev's): $\lang F^T\rang = 0,$ thus avoiding a potentially dangerous
source of flavor changing neutral currents (FCNC).
\item The constraint of vanishing (or nearly so) vacuum energy leads
to a variety of mass hierarchies that involve the $\beta$-function
coefficient of the condensing gauge group.
\end{itemize}

These results were obtained at tree level in the effective
supergravity theory for gaugino condensation, which includes the
quantum corrections in the strongly coupled gauge sector whose
elementary degrees of freedom have been integrated out, as well as the
four dimensional Green-Schwarz (GS) terms needed at the quantum level
to cancel field theory anomalies.  In addition, the logarithmically
divergent and finite (``anomaly mediated''~\cite{anom}) one-loop
corrections to soft supersymmetry-breaking parameters have been
extensively studied~\cite{gnw,gns,bgn}. These analyses did
not include quadratically divergent loop corrections that are for the
most part corrections to terms in the tree Lagrangian, and
are suppressed by the loop expansion parameter
\beq \epsilon = 1/16\pi^2.\label{ep}\eeq
However, since some of these terms have coefficients proportional to
the number of fields in the effective supergravity theory, it has been
argued that they may not be negligible.  In particular, their
contributions to the cosmological constant~\cite{ckn} and to flavor
changing neutral currents~\cite{clm} have been emphasized. 
Both are important for the phenomenology of the above
condensation models; thus we need to revisit~\cite{gnq} their effects.

\section{Modular invariant gaugino condensation}
One starts above the (reduced) Planck scale $m_P$ with the heterotic
string theory in 10 dimensions.  Just below the string scale $\mu_s =
g_s m_P$, where $g_s$ is the gauge coupling at the string scale,
physics is described by $N=1$ modular invariant supergravity in four
dimensions, where here modular invariance refers to T-duality under
which the K\"ahler moduli $T$ transform as
\beq T\to\frac{a T - i b T}{i c T + d},\qquad a d - b c = 1, \qquad
a,b,d,c\in \mathcal{Z}.\eeq
Modular invariance -- and in many compactifications~\cite{jg} a \u1\,
gauge group factor called \ux\, -- is broken by anomalies at the
quantum level of the effective field theory, and the symmetry is
restored by an appropriate combination of threshold
effects~\cite{thresh} and four dimensional GS
term(s)~\cite{gs1,gs2}.  The precise form of these loop effects in the
Yang-Mills sector of the effective supergravity theory have been
determined by matching the string and field theory amplitudes at the
quantum level~\cite{gt}.

If an anomalous \u1\, is present, the corresponding GS term leads to a
Fayet-Illiopoulos (FI) D-term in the effective Lagrangian~\cite{gs2}
and some \u1-charged scalars $\phi^A$ acquire \vev's at a scale
$\mu_D$ one or two orders of magnitude below the Planck scale such
that the overall D-terms vanish:
\beq \lang\frac{1}{\ell(s,\s)}\sum_A q^a_A(t +
\t)^{n_A}|\phi^A|^2\rang = \half\delta_X\delta_{Xa},\label{dterm}\eeq
where $\delta_X\ell$ is the
coefficient of the FI term, $n^A$ is the modular weight of $\phi^A$,
$q^a_A$ is its charge under the gauge group factor \ua, and $t,s$ are
the scalar components of the K\"ahler moduli and dilaton chiral
superfields $T,S$. The function $\ell(s + \s)$ is the dilaton field in
the dual, linear supermultiplet formulation; in the classical limit
$\ell = (s + \s)^{-1}$.  The combination of fields that gets a vacuum
value is modular invariant.  Thus modular invariance, as well as local
supersymmetry, is unbroken at this scale, and the moduli fields $s,t$
remain undetermined~\cite{ggu}. The $\phi^A$ vacuum is generically
characterized by a high degree of further degeneracy~\cite{bdfs,ggd}
that may lead to problems for cosmology.

At a lower scale $\mu_c$, a gauge group $\mathcal{G}_c$ in the hidden
sector becomes strongly coupled, and gauginos as well
$\mathcal{G}_c$-charged matter condense.  The potential generated for
the moduli is T-duality invariant and the K\"ahler moduli $T$ are
stabilized at self-dual points with $\lang F^T\rang = 0$, while $\lang
F^S\rang\ne 0$, so that, in the absence of an anomalous \u1,
supersymmetry breaking is dilaton mediated~\cite{bgw}. In the presence
of an anomalous \u1, \vev's of D-terms are generically generated as
well and tend to dominate supersymmetry breaking; these may be
problematic for phenomenology.  On the plus side, at least some of the
degeneracy of the $\phi^A$ vacuum is lifted by $\phi^A$ couplings to
the condensates~\cite{ggm}.
  
To briefly summarize the phenomenology of these models, the condition
of vanishing vacuum energy introduces the $\beta$-function coefficient
of the condensing gauge group $\mathcal{G}_c$
into the supersymmetry breaking parameters in
such a way as to generate a variety of mass hierarchies.  Defining
\beq b_c = \frac{1}{16\pi^2}\(3C^c - C^c_M\),\eeq
where $C^c(C^c_M)$ is the adjoint (matter) quadratic Casimir for
$\mathcal{G}_c$, in the absence of an anomalous \u1\, one has at the
condensation scale~\cite{bgw} (one can also have $m_0\sim m_T\gg\mG$ if
gauge-charged matter couples to the GS term)
\bea m_0 &=& \mG,\qquad m^a_{\half} =
\frac{4b^2_c}{9}g_a(\mu_c)\mG,\nonumber\\ m_T
&\approx&\frac{b}{b_c}\mG, \qquad m_S\sim b_c^{-2}\mG,\qquad m_a =0.
\label{masses}\eea
where $m_{0,\half,\frac{3}{2}}$ refer to observable sector scalars and
gauginos, and the gravitino, respectively; $m_{T,S,a}$ are the
K\"ahler moduli, dilaton and universal axion masses.  The expression
for $m_T$ assumes $b\gg b_c$, where $b$ is the $\beta$-function
coefficient appearing in the modular invariance restoring GS
term~\cite{gs1}. For example in the absence of Wilson lines, $b= b_{E8}
\approx .57$, and viable scenarios for electroweak symmetry
breaking~\cite{gnph} and for neutralinos as dark matter~\cite{bhn}
require $b_c\approx .05-.06$.  These numbers give desirably large
moduli and dilaton masses, while the scalar/gaugino mass ratio is
perhaps uncomfortably large, but no worse than in many other models.

When Wilson lines are present the condition $b\gg b_c$ may not hold;
for example $b_c = b$ in a $Z_3$ compactification~\cite{fiqs} with an
$SO(10)$ hidden sector gauge group; this would give vanishing T-moduli
masses in the above class of models.  However when an anomalous \u1 is
present, the T-moduli couplings to the condensates are modified,
giving additional contributions to their masses, and a hierarchy with
respect to the gravitino mass can still be maintained~\cite{ggm}. In
this scenario the gaugino, dilaton and axion masses are determined
only by the dilaton potential, as before.  A stable vacuum with a
positive metric for the dilaton is most easily achieved in a
``minimal'' class of models in which the number of Standard Model (SM)
gauge singlets that get \vev's at the scale $\mu_D$ is equal to the
number $m$ of broken \u1's (in which case there are no massless
``D-moduli''~\cite{ggd} associated with the degeneracy of the
\u1-charged $\phi^A$ vacuum), or $N$ replicas of these with identical
\u1\, charges [yielding $(N-1)m$ D-moduli].  In this case the gaugino
and dilaton masses are unchanged from (\ref{masses}) (the axion mass
always vanishes).  The most significant change from the above scenario
is a D-term contribution to scalar squared masses $m_0^2$ that is
proportional to their \u1\, charges. At weak coupling, and neglecting
nonperturbative effects, this term dominates the one in (\ref{masses})
by a factor $b_c^{-2}\gg1$, and is not positive semi-definite.  Thus
unless SM particles are uncharged under the broken \u1's (or have
charges that, in a well-defined sense~\cite{ggm}, are orthogonal to
those of the $\phi^A$ with large \vev's), these models are seriously
challenged by the SM data: a very high scalar/gaugino mass ratio for
positive $m_0^2$, and the danger of color and electromagnetic charge
breaking if $m_0^2<0$.

\section{Quadratically divergent corrections}
When local supersymmetry is broken, there is a quadratically
divergent one-loop contribution to the vacuum energy~\cite{zum}
\beq \lang V_{\rm 1-loop} \rang \ni \frac{\Lambda^2}{32\pi^2}
\lang\STr \mathcal{M}^2\rang,\eeq
where $\mathcal{M}$ is the field-dependent mass matrix, and the
gravitino contribution is gauge dependent.  For example in minimal
supergravity~\cite{can} with $N_{\chi}$ chiral and $N_G$
Yang-Mills  superfields, one obtains, using the gravitino gauge fixing
procedure of Ref~\cite{gjs}.
\begin{equation} 
\lang \delta V_{\rm 1-loop} \rang
\ni\frac{\Lambda^2}{16\pi^2}\(N_{\chi} m_0^2 -N_G\mg^2 + 2\mG^2\) .
\label{STr} \end{equation}
In the MSSM we have $N_{\chi} = 49$ and $N_G = 12$. The much larger
field content of a typical $Z_3$ orbifold
compactification~\cite{cmm,jg} of the $E_8\otimes E_8$ heterotic string
has $N_\chi\gappeq 300$ and $N_G \lappeq 65$, suggesting~\cite{ckn}
that this contribution to the vacuum energy is always positive.

However, in order to maintain manifest supersymmetry, a supersymmetric
regularization of ultraviolet divergences must be used.  Pauli-Villars
(PV) regularization~\cite{pv} meets this criterion.  The regulation of
quadratic divergences requires {\it a priori} two subtractions; in the
context of PV regularization, the number $S$ of subtractions is the
number of PV fields for each light field. Once the divergences are
regulated ({\it i.e.} eliminated), we are left with the replacement
\begin{equation}
\Lambda^2 \STr\mathcal{M}^2 \to \STr \mu^2 \mathcal{M}^2 \ln(\mu^2)
\eta_S, \qquad \eta_S = \sum_{q=1}^S \eta_q \lambda_q \ln \lambda_q,
\label{subtract} \end{equation}
where $\mu$ represents the scale of new physics, and the parameter
$\eta_S$ reflects the uncertainty in the threshold for the onset of
this new physics. The squared PV mass of the chiral supermultiplet
$\Phi^q$ is $\lambda_q\mu^2$ (so $\lambda_q>0$), and $\eta_q = \pm1$
is the corresponding PV signature.  The sign of the effective cut-off
is determined by the sign of $\eta_S$, which is positive definite
only\footnote{See appendix C of \cite{bgsig}. and the discussion
in \cite{gnq}.} if $S\le3$. Cancellation of all the ultraviolet
divergences of a general supergravity theory requires~\cite{bgpr} at
least 5 PV chiral multiplets for every light chiral multiplet and even
more PV supermultiplets to regulate gauge loops.  Therefore one cannot
assume that the effective cut-offs are all positive.

Including the full (cut-off) one-loop quadratically divergent
contribution gives the effective bosonic Lagrangian
\bea e^{-1}\L(\Lambda) &=& e^{-1}\L_{\rm tree} +
\epsilon\Lambda^2\frac{r}{4}\(N_\chi- 7 - N_G\)\nonumber \\ & & +
\epsilon\Lambda^2\(5V + \mG^2 - 2K_{i\m} \D_\mu z^i\D^\mu\z^{\m} -
\D\)\nonumber \\ & & - \frac{\epsilon\Lambda^2}{\re s} \D_aD_i(T^az)^i
+ \epsilon\Lambda^2 R_{i\m}\(F^i\bar F^{\m} + \D_\nu
z^i\D^\mu\z^{\m}\) \nonumber \\ & & - \epsilon\Lambda^2 N_\chi\(V +
\mG^2 - \D\) + \epsilon\Lambda^2 N_G\(\frac{\pp_\mu s\pp^\mu\s}{(s +
\s)^2} + \mg^2\),\eea
where $\mG,\mg$ are now understood to be field-dependent, 
\beq \D_a = K_i(T^a z)^i, \qquad F^i = - e^{K/2}K^{i\m}\bar W_{\m},
\eeq
are the usual auxiliary fields, the tree level potential is
\beq V_{\rm tree} = V = D + K_{i\m}F^i\bar F^{\m} - 3\mG^2, \qquad
D = \frac{D^a\D_a}{s + \s},\label{potential}\eeq
and $R_{i\m}$ is the K\"ahler Ricci tensor.  After a Weyl
transformation to restore the Einstein term to canonical form, we
obtain
\bea e^{-1}\L(\Lambda) &=& e^{-1}\L_{\rm tree}(g_R) +
\epsilon\Lambda^2\(\mG^2 - 2V + \frac{3}{2}K_{i\m}\D_\mu
z^i\D^\mu\z^{\m} - \D\)\nonumber \\ & & - \frac{\epsilon\Lambda^2}{\re
  s} \D_aD_i(T^az)^i + \epsilon\Lambda^2 R_{i\m}\(F^i\bar F^{\m} +
\D_\nu z^i\D^\mu\z^{\m}\) \nonumber \\ & & - \epsilon\Lambda^2
N_\chi\(\mG^2 - \D + \half K_{i\m}\D_\mu z^i\D^\mu\z^{\m}\) \nonumber
\\ & & + \epsilon\Lambda^2 N_G\(\frac{\pp_\mu s\pp^\mu\s}{(s + \s)^2}
+ \mg^2 - V + \half K_{i\m}\D_\mu z^i\D^\mu\z^{\m}\),\label{lco}\eea
where $g_R$ is the one-loop renormalized metric.  The Lagrangian
(\ref{lco}) does not respect supersymmetry.  With a supersymmetric
PV regularization, PV masses arise from quadratic couplings in
the superpotential 
\beq W_{PV} \ni \mu_{I J}(Z^k)Z^I_{PV}Z^J_{PV},\qquad
\left.Z^k \right|= z^k.\label{wpv}\eeq
Then the squared cut-off in (\ref{lco}) is replaced by suitably
weighted linear combinations of PV squared masses
\beq \Lambda^2 \to (M^2)^I_J = e^K K^{I\bar K}(z)K^{\bar L M}(z)
\bar\mu_{\bar K\bar L}(\z)\mu_{M J}(z)\eeq
that are generally field-dependent.  Moreover, the couplings
(\ref{wpv}) induce additional terms proportional to $M^2$ that
cannot be obtained by a straight cut-off procedure.  The resulting
effective Lagrangian takes the form~\cite{pv}
\beq \L^1_{eff} = \L_{\rm tree}(g,K) + \L_{\rm 1-loop} = \L_{\rm
tree}(g_R,K_R) + O(\epsilon\ln\Lambda_{eff}^2) + O(\epsilon^2),\eeq
where 
\beq K_R = K + \Delta K \eeq
is the renormalized superpotential.  The action obtained in this
way is only perturbatively supersymmetric:
\beq \delta S^1_{eff}  = \int d^4x \delta\L^1_{eff} = O(\epsilon^2).\eeq
Writing 
\beq \Delta K = \frac{\epsilon}{2}\[N\Lambda_\chi^2 -
4N_G\Lambda_G^2 +O(1)\Lambda^2_{\rm grav}\]
+ O(\epsilon\ln\Lambda_{eff}^2) + O(\epsilon^2),\label{delk} \eeq
where $\Lambda_{\chi,G,{\rm grav}}$ are the effective cut-offs for chiral,
gauge and gravity loops, and $\Lambda_{eff}$ is a generic effective
cutoff, if $N_\chi,N_G\sim \epsilon^{-1}$, we must retain the full
effective Lagrangian as derived from $K_R$.  This amounts to
resuming the leading terms in $\epsilon N\Lambda_{eff}^2$, with the
result, as dictated by supersymmetry, just a correction to the
K\"ahler potential.  I will discuss the consequences of this correction
in the following sections.

\section{The vacuum energy}
Consider first the possibility that we can choose the $Z^k$-dependence
of the PV K\"ahler potential and superpotential such that the
effective cutoffs are constant.  For example, one needs PV superfields
$Z^I_{PV}$ with the same K\"ahler metric as the light superfields
$Z^i$: $K^Z_{I\bar M} = K_{i\m}.$ If we introduce superfields $Y_I$ with
K\"ahler metric: $K_Y^{I\bar M} = e^{-K}K^{-1}_{i\m} = e^{-K}K^{i\m},$
the superpotential coupling
\beq W_{PV} = \mu Z^I Y_I\eeq
yields a constant squared mass $M^2 = \mu^2$ if $\mu$ is constant,
and the quantum corrected potential just reads
\beq V_{eff} = \D + e^{\Delta K}\(F^i K_{i\m} F^{\m} - 3\mG^2\)_{\rm
tree} + O(\epsilon\ln\Lambda_{eff}^2).\label{const}\eeq
If supersymmetry breaking is F-term induced: $\lang\D\rang=0$, the
tree level condition $\lang F^i K_{i\m} F^{\m} = 3\mG^2\rang$ for
vanishing vacuum energy is unmodified by these quantum corrections.

However not all PV masses can be chosen to be constant because of the
anomaly associated with K\"ahler transformations $K(Z,\bar Z)\to
K(Z,\bar Z) + F(Z) + \bar F(\bar Z)$ that leave the classical
Lagrangian invariant.  In the presence, for example, of an anomalous
\ux, with generator $T_X$, there is a quadratically divergent term
proportional to $\Tr T_X\Lambda^2$ that cannot be canceled by
\ux-invariant PV mass terms, since the contribution to $\Tr T_X$ from
each pair in the invariant superpotential cancels. As a consequence,
there must be some PV masses $\propto e^{a V_X}$, where $V_X$ is the
\ux\, vector superfield.  Similarly, in the presence of a K\"ahler
anomaly there is a term
\beq\L_{\rm 1-loop}\ni c\epsilon K_{i\m}\D_\mu
z^i\D^\mu\z^{\m}\Lambda^2,\label{kanom}\eeq
that cannot be canceled unless some PV superfields have masses
$M^2_{PV}\propto e^{\alpha K}$.  In addition, PV regulation of
the gauge + dilaton sector requires some PV masses proportional
to the field-dependent string-scale gauge coupling constant:
$M^2_{PV}\propto g^s(s,\s) = 2(s + \s)^{-1}.$  

What are the effects of this field-dependence on the condensation
models described above?  In order to implement the correct Bianchi
identity for the gaugino condensate composite superfield -- as well as
the GS anomaly cancellation mechanism -- the linear
multiplet formulation for the dilaton was used in the construction of
the effective Lagrangians for these models.  The results have been
recast~\cite{bgn} in the more familiar language of the chiral multiplet
formalism, so that the effective tree-level
potential below the scale of condensation takes the standard form
(\ref{potential}) with
\beq \mG = \frac{3}{2} b_c u, \qquad F^S = - \frac{1}{4}K_{S\S}^{-1}\(1 -
\frac{2}{3}b_c K_S\)\bar u, \eeq
where $u$ is the vacuum value of the condensate.
The modular invariance of these models assures that the
moduli $T$ are stabilized at self-dual points with vanishing
auxiliary fields: $\lang F^{T}\rang =0$. Supersymmetry breaking
is dilaton-dominated and the condition for vanishing vacuum energy
at tree level in the effective theory is
\beq \lang V_{eff}\rang = \lang K_{S\S}|F^S|^2 -
3\mG^2\rang = 0, \qquad \lang K^{-1}_{S\S}\rang =
\frac{4b_c^2}{3(1 - \frac{2}{3}\lang K_S\rang
b_c)^2},\label{vacen}\eeq
Classically,
\beq -2\lang K_S\rang = 2\lang K^{\half}_{S\S}\rang = 2\lang(s +
\s)^{-1}\rang = g_s^2 \approx \half. \eeq
with the approximate value of $g_s$ inferred from low energy
data.  The model is phenomenologically~\cite{gnph} and
cosmologically~\cite{bhn} viable if $b_c\approx .05$--$.06$, so it is
clear that (\ref{vacen}) cannot be satisfied without a modification of
the K\"ahler potential for the dilaton; one approach~\cite{bgw1} has
been to invoke nonperturbative string~\cite{shank} and/or QFT~\cite{bd1}
corrections to the dilaton K\"ahler potential.  Specifically we
require
\beq K^{-1}_{s\s}\ll \left.  K^{-1}_{s\s}\right|_{\rm classical}.\eeq
Avoiding dangerously large D-term contributions to scalar masses in
the presence of an anomalous \u1\, may further require~\cite{ggm}
\beq - K_s\gg \left. - K_s\right|_{\rm classical},\eeq
suggesting that weak coupling may not be viable~\cite{bdhw}.

To examine the effects of quadratically divergent perturbative
corrections, I assume a form of the superpotential suggested by
the discussion following (\ref{kanom}):
\bea K^R &=& K + \half\epsilon c_\chi f(T,\T)e^K - \frac{4\epsilon
c_G}{ S + \S - V_{GS}}\nonumber \\ &=& K + \half\epsilon c_\chi
f(T,\T)e^K - 2\epsilon c_G g_s^2(Z,\bar Z),\label{newk}\eea
where $V_{GS}(T + \T)$ is the Green-Schwarz term (which slightly
redefines the string-scale coupling $g_s$ at the quantum level), and
$f(T,\T)$ assures modular invariance of the second term.  The
loop-induced terms may not be negligible if $c_\chi,c_G\sim
N_\chi,N_G$, respectively. Since the theory is still modular invariant
we expect that the moduli are still stabilized at self-dual points,
where the additional contributions to $F^{T}$ induced by these quantum
corrections vanish. Setting the T-moduli at their \vev's and the
matter fields to zero, and defining
\beq K = k(2g_s^{-2}) + G(T+\T), \qquad \tc_\chi = \lang{f(t,\t)e^G}\rang
c_\chi,\eeq
the renormalized K\"ahler potential and its $S$-derivatives read
\bea K^R &=& k + \frac{\epsilon}{2}\(\tc_\chi e^k - 4c_G g^2_s\),
\qquad K^R_S = K_S\(1 + \frac{\epsilon}{2}\tc_\chi e^k\) + \epsilon
c_G g_s^4,\nonumber \\ K^R_{S\S} &=& K_{S\S}\(1 +
\frac{\epsilon}{2}\tc_\chi e^k\) + \frac{\epsilon}{2}\(K^2_S\tc_\chi
e^k - c_G g_s^8\).\eea
The condition for vanishing vacuum energy is now given by
(\ref{vacen}) with $K\to K^R$, and the relevant parameter for
phenomenology is now the \vev\, of $1/K_{S\S}^R$, which remains
strongly suppressed with respect to its classical value since $K^R_S$
is negative semi-definite.\footnote{The relation $\ell = - K^R_S$
holds at any given order in perturbation theory, where $\ell$ is the
dilaton of the dual linear multiplet formalism.}  Therefore the
salient phenomenological features of the condensation models are
essentially unaffected by these quantum corrections.

However these corrections could lessen the need to invoke
large nonperturbative effects.  A large {\it negative} value of $c_G$
or a large {\it positive} value of $\tc_\chi$ would increase $-K^R_S$
and decrease $1/K^R_{S\S}$ for fixed $g^2_s\approx 1/2$, which is the
desired effect.  One can reasonably assume that $|c_G|\le N_G \le 65
\sim.4\epsilon^{-1}$ in typical orbifold compactifications~\cite{jg},
so a significant effect cannot be obtained from the second term in
(\ref{newk}).  On the other hand $N\epsilon\sim 2$ for typical
orbifolds. Quite generally the function $f(t,\t)$ is of order one at a
self-dual point, so if $\tc_\chi \sim N$, $e^k\sim 1$, it might be
possible to reinterpret part of the needed modification of the dilaton
K\"ahler potential in terms of perturbative quantum corrections.

\section{Flavor Changing Neutral Currents}
To address the question of what constraints are needed to avoid
experimentally excluded FCNC effects, we first note that the tree
potential of an effective supergravity theory includes a term
\beq V_{\rm tree}\ni e^K K_i
K_{\jbar}K^{i\jbar}|W|^2.\label{vkahl}\eeq
The observed suppression of FCNC effects constrains the K\"ahler
potential; to a high degree of accuracy we require that
\beq K_i K_{\jbar}K^{i\jbar} \not{\hspace{-.03in}\ni} \lang f(X,\bar
X)\rang \phi^A_f\bar\phi^{\bar A}_{f'\ne f},\eeq
where $f,f'$ are flavor indices, $A$ is a gauge index, $\phi^A_f$ is
any standard model squark or slepton, and $X$ is a singlet of the
Standard Model gauge group. For example, in the no-scale models that
characterize the untwisted sector of orbifold compactifications, we
have
\beq K_i K_{\jbar}K^{i\jbar} = 3 + K_S K_{\bar S} K^{S\bar S},\eeq
which is safe, since $K_S$ is a function only of the dilaton.  The
twisted sector K\"ahler potential is known only to quadratic order:
\beq K_T = \sum_A(T + \bar T)^{n_A}|\Phi^A_T|^2 + O(\Phi^3),
\label{tw}\eeq
which is flavor diagonal and also safe.  The higher order terms in
(\ref{tw}) could be problematic if some $\phi^A = X^A$ have large
\vev's ({\it i.e.} within a few orders of magnitude of the Planck
scale).  Thus phenomenology requires that we forbid couplings of the
form $\phi^A_f\phi^{\bar A}_{f'\ne f}|\phi_{f''}^{A'}|^2X^{B_1}\cdots
X^{B_n}$, $n\le N$, where $N$ is chosen sufficiently large to make the
contribution $\lang X^{B_1}\cdots X^{B_n}\rang$ to the scalar mass
matrix negligible.  The quadratically divergent one-loop corrections
generate a term
\beq V_{\rm 1-loop}\ni e^K K_i
K_{\jbar}R^{i\jbar}|W|^2, \qquad R^{i\jbar} = K^{i\bar k}R_{\bar k l}
K^{k\jbar}.\label{vricc}\eeq
where $R_{i\jbar}$ is the K\"ahler Ricci tensor.  The contribution
(\ref{vricc}) simply reflects the fact that the leading divergent
contribution in a nonlinear sigma model is a correction to the
K\"ahler metric proportional to the Ricci tensor (whence, {\it e.g.},
the requisite Ricci flatness of two dimensional conformal field
theories).  Since the Ricci tensor involves a sum of K\"ahler Riemann
tensor elements over all chiral degrees of freedom, a large, order
$N_\chi$, coefficient may be generated~\cite{clm}. For example, for an
untwisted sector $U$ with three untwisted moduli $T^n$ and K\"ahler
potential
\beq K^U = \sum_{n=1}^3K^n = - \sum_{n=1}^3\ln(T^n + \bar T^{\bar n} -
\sum_{A=1}^{N_n}|\Phi^A_n|^2),\label{nskp} \eeq
we get
\beq R^n_{i\jbar} = (N_n + 2)K^n_{i\jbar}.\label{runtw}\eeq 
While this contribution is clearly safe, since the Ricci tensor is
proportional to the K\"ahler potential, the condition that the tree
potential be FCNC safe does not by itself ensure that (\ref{vricc}) is
safe in general.  For this we require in addition the absence of
K\"ahler potential terms of the form $\phi^A_f\bar\phi_{f'\ne f}^{\bar
A}|\phi_{f''}^{A'}|^4(X^B)^{n\le N}$.  On the other hand, if the
K\"ahler metric is FCNC safe due to a {\it symmetry}, the same
symmetry will protect the Ricci tensor from generating FCNC.

For example, the scalar metric $g_{i j}$ for the effective pion
Lagrangian is dictated by chiral $SU(2)_L\otimes SU(2)_R$; there is a
unique form of the two-derivative coupling: 
\beq g_{i j} = \delta_{i j} + \frac{\pi_i\pi_j}{ v^2 - \pi^2},\eeq
for a particular choice of field variables.  Preservation of this
symmetry at the one-loop level assures that $R_{i j}\propto g_{i j}$.
Similarly, the kinetic term derived from the K\"ahler potential
(\ref{nskp}) possesses an $\prod_{n=1}^3SU(N_n + 1,1)$ symmetry that
is much larger than the $SL(2,R)$ (or possibly$[SL(2,R)]^3$) T-duality
symmetry of the full Lagrangian, and we obtain the result
(\ref{runtw}).  More generally, in effective supergravity from string
compactifications there are a number of selection rules and/or
symmetries that forbid superpotential couplings that are allowed by
gauge invariance and the T-duality invariance group (see for
example~\cite{fiqs}).  The K\"ahler potential has not been
investigated in similar detail, but {\it a priori} one would expect an
analogous pattern. In the absence of input from string theory one can
work backwards and study~\cite{gkn} the constraints imposed by
phenomenology.

Another handle on this issue is the requirement of the full
cancellation of \ux\, and modular anomalies in the fully regulated
effective supergravity theory.  This requirement may
restrict~\cite{bgpr} the K\"ahler potential couplings of both the
twisted sector and of the regulator PV fields that parameterize
Planck scale physics.

\section{Conclusions}
I have outlined some of the promises as well as the problems of string
phenomenology in models with supersymmetry broken by gaugino
condensation.  The issue of quadratic divergences in the effective
supergravity theory was rephrased as a renormalization of the K\"ahler
potential, with little impact on the phenomenology of these models,
except for a possible reinterpretation of ``string nonperturbative
effects'' in terms of perturbative contributions to the renormalized
K\"ahler potential.  I argued that FCNC bounds may constrain the
twisted sector K\"ahler potential as well as that of PV fields. Such
constraints would provide a ``bottom up'' probe of Planck scale
physics.  Conversely, the PV masses, that are determined in large part
by the PV K\"ahler potential, play an important role in determining
scalar masses if they are dominated by one-loop corrections~\cite{gns}.
Therefore any constraints on the K\"ahler potential from string theory
calculations and/or the requirement of anomaly cancellation could
provide a ``top down'' contribution to collider physics.

\section*{Acknowledgments}
I wish to thank my collaborators on the work described here.  This
work was supported in part by the Director, Office of Energy Research,
Office of High Energy and Nuclear Physics, Division of High Energy
Physics of the U.S. Department of Energy under Contract
DE-AC03-76SF00098 and in part by the National Science Foundation under
grants PHY--0098840.


\end{document}